\documentclass[prb,twocolumn,,amsmath,amssymb,floatfix]{revtex4}
\usepackage{graphicx}
\usepackage{bm}

\newcommand{\be}{\begin{equation}}
\newcommand{\ee}{\end{equation}}

\def \be{\begin{equation}}
\def \ee{\end{equation}}
\def \ba{\begin{array}}
\def \ea{\end{array}}
\def \beq{\begin{eqnarray}}
\def \eeq{\end{eqnarray}}

\begin{document}

\title{Linear response theory for a pair of coupled one-dimensional condensates of interacting atoms}

\author{Vladimir Gritsev$^1$, Anatoli Polkovnikov$^2$, Eugene Demler$^1$ }
\affiliation {$^1$Department of Physics, Harvard University, Cambridge, MA 02138\\
 $^2$ Department of Physics, Boston University, Boston, MA 02215}

\begin{abstract}
We use quantum sine-Gordon model to describe the low energy dynamics
of a pair of coupled one-dimensional condensates of interacting
atoms. We show that the nontrivial excitation spectrum of the
quantum sine-Gordon model, which includes soliton and breather
excitations, can be observed in experiments with time-dependent
modulation of the tunneling amplitude, potential difference between
condensates, or phase of tunneling amplitude. We use the form-factor
approach to compute structure factors corresponding to all three
types of perturbations.
\end{abstract}

\date{\today}

\maketitle

\section{Introduction}

When discussing two dimensional classical or one dimensional quantum
systems, no model has a wider range of applications than the
sine-Gordon model. The sine-Gordon model (SG) was originally studied
in the context of high-energy physics\cite{Col,DHN} and later became
a prototypical model of low-dimensional condensed matter systems. In
statistical physics it has been successfully applied to describe the
Kosterlitz-Thouless transition in two dimensional superconductors
and superfluids \cite{NH,FishGr} as well as commensurate to
incommensurate transitions\cite{JN} on surfaces. The quantum
sine-Gordon model was used to describe the superfluid to Mott
transition of bosons in one dimension in such systems as Cooper
pairs in Josephson junctions arrays\cite{Kar} and cold atoms in
optical lattices\cite{Fishers}. Sine-Gordon model also provides a
useful framework for understanding properties of one dimensional
spin systems with easy axis/plane anisotropies or in the presence of
magnetic field (see [\onlinecite{GNT}] for review). Interacting
electron systems in one dimension often exhibit phase transitions
between states with short range and power law correlations. Such
transitions are also expected to be in the universality class of the
quantum KT transition described by the quantum sine Gordon
model\cite{GNT}. Large classes of boundary-related phenomena
\cite{GZ}, including the problem of an impurity in an interacting
electron gas (the so-called quantum impurity model)(see
[\onlinecite{Sal}] for review), Josephson junctions with
dissipation\cite{Sal}, Kondo-like models\cite{FLS}, single electron
transport in ultrasmall tunnel junctions have also been discussed in
the framework of the boundary sine-Gordon model. Quantum sine-Gordon
model provides a universal description for such wide range of
systems because it is the simplest model with a gapped spectrum and
relativistic low energy dispersion.

One approach to understanding the sine-Gordon model is based on the
renormalization group analysis (see Refs.~[\onlinecite{GNT}] and
[\onlinecite{GG}] for the review). This method has been successfully
applied to discuss thermodynamic and transport properties of two
dimensional superfluids and superconductors~\cite{NH}.

Another more detailed approach is based on the integrability of the
model and existence of the complete exact solution~\cite{Fad,ZZ}.
From this analysis we know that the excitation spectrum of the model
consists of solitons, antisolitons and their bound states-breathers.
The number and properties of different breathers is controlled by
the interaction strength (related to the parameter $\gamma$
introduced in section IB below).

While theoretical understanding of the quantum SG model is quite
advanced, very few experimental studies  have been done that
addressed dynamics of this system. Several experimental studies on
one-dimensional magnets interpreted resonances in neutron scattering
and ESR experiments as breather type excitations \cite{AO,EFH,Zv}.
All these examples include spin-1/2 magnetic chains perturbed by an
effective $g$-tensor and the Dzyaloshinskii-Morya interaction  or by
a staggered field in copper benzoate \cite{DMg,As} and
dimethylsulfoxide \cite{KCBRQ} and/or by strong magnetic field, like
in copper pyrimidine dinitrate\cite{Zv}. Apart from these
measurements we do not have experimental evidences to verify our
understanding of the spectrum of the quantum SG model. Considering
the importance of the sine-Gordon model for understanding one
dimensional quantum systems it is of great interest to find new,
more direct approaches for experimental investigation of this
fundamental model.

\begin{figure}[hb]
\includegraphics[width=8.5cm,angle=0]{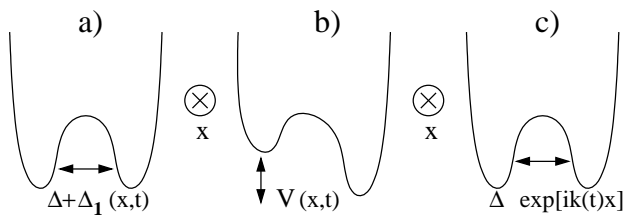}
\caption{ Three types of experimental setups proposed to study
linear response in two coupled condensates. The figure shows a
projection of the confinement potential. In setup a) the tunneling
amplitude is modulated: $\Delta(x,t)=\Delta+\Delta_{1}(x,t)$ (see
Eqs.~(\ref{tun}), (\ref{sgham})); setup b) corresponds to the space-
time-dependent population imbalance accessed via the change $V(t)$
of the confinement depth in one of the well; in experiment of type
c) the space-time modulation comes from winding of the phase of
single particle tunneling, $\Delta(x,t)=\Delta\exp(i k(t)x)$.
$\Delta$ is parameter defined in Eq.~(\ref{tun}). The $x$-direction
coincides with the longitudinal direction of condensates and is out
of plane of the figure.}
\label{exp-pic}
\end{figure}

In this paper we show that a pair of coupled one-dimensional
condensates provides realization of the quantum Sine-Gordon model.
Such systems were recently realized in cold atoms in microtraps,
where the RF potential controls the tunneling between the
condensates~\cite{Sch,Sch2,jo}. The advantage of using cold atoms to
realize the quantum SG model is that cold atom systems are highly
tunable. For example in Refs.~[\onlinecite{KWW}-\onlinecite{KWW2}]
it was demonstrated that the interaction and thus the Luttinger
parameter describing one-dimensional condensates can be varied in a
very wide range. Similarly the tunneling between the two
condensates, which controls the gap in the sine-Gordon theory (see
details below), can be tuned to a high precision. Coupled
condensates can also be realized in optical lattices using
superlattice potentials~\cite{nist,Sch}.

One possibility to probe the excitation spectrum of the quantum
sine-Gordon model is to study the response of the system to small
periodic modulations of various parameters of the model. In this
paper we discuss three types of such modulation experiments:
modulation of the tunneling amplitude, modulation of the potential
difference between the condensates, and modulation of the phase of
the tunneling amplitude (see Fig.~\ref{exp-pic}). By observing the
frequency dependence of the absorbed energy in various modulation
experiments, one can measure structure factors associated with
appropriate perturbations. Here we theoretically compute structure
factors for all three types of modulations (see Figs.~\ref{cos1},
\ref{cos2}, \ref{dphi2} and \ref{dphi3}). As we demonstrate below,
different types of modulations expose different parts of the quantum
sine-Gordon spectrum and provide complimentary information about the
system. We note that the calculations of structure factors for
several types of operators in the quantum sine-Gordon model was done
in earlier papers in the context of condensed matter
systems~\cite{CET,EGJ,ET}, mainly to describe properties of Mott
insulating states. These results were summarized in a recent
review~\cite{EK}. Somewhat similar analysis has been done in the
context of the boundary sine-Gordon model (see
Ref.~[\onlinecite{Sal}] for the review). In this paper we apply and
extend this earlier analysis.

We note in passing that in a separate publication we discussed a
different possibility to study the structure of the sine-Gordon
model using quench experiments~\cite{quench}. There we showed that
the power spectrum of the phase oscillations after a sudden split of
a single condensate into two contains detailed the information about
various excitations in the sine-Gordon model. The quench and
modulation experiments are thus complimentary ways to extract this
information.

This article is organized as follows. First we give an effective
description of the system in Sec.~\ref{effsec}. Then we briefly
summarize general facts about the quantum sine-Gordon model in
Sec.~\ref{sgfacts}. In Sec.~\ref{sgsec} we outline the explicit
construction of the sine-Gordon form-factors. In
Sec.~\ref{linrespsec} we evaluate and analyze the structure factors
for different kind of perturbation. Finally in Sec.~\ref{sumsec} we
discuss experimental implications and extensions of our work and
summarize the results.

\section{Microscopic Model.}

\subsection{Effective model of two coupled condensates\label{effsec}}

We consider a system of two coupled interacting one-dimensional
condensates. If interactions between atoms are short ranged, the
Hamiltonian providing a microscopic description of the system is
\beq\label{twocond}
&&H=\sum_{j=1,2}\int d x \left[-{\hbar^2\over
2m}\partial_{x}\Psi_{j}^{\dagger}(x)\partial_{x}\Psi_{j}(x)+g
n_{j}^{2}(x)\right]\nonumber\\
&&~~~~~~~~~-t_{\perp}\int dx \left(\Psi_{1}^{\dagger}(x)
\Psi_{2}(x)+\Psi_{2}^{\dagger}(x) \Psi_{1}(x)\right),
\eeq
where $n_j(x)=\Psi_{j}^\dagger (x) \Psi_{j}(x)$. Here $t_\perp$ is
the coupling, which characterizes the tunneling strength between the
two systems, and $g$ is the interaction strength related to the 3D
scattering length $a_{3D}$ and the transverse confinement length
$a_\perp=\sqrt{\hbar/m\omega_\perp}$ via:~\cite{O}
\be\label{golsh}
g\approx {4\hbar^2\over m a_\perp} {a_{3D}\over
a_\perp}\left(1-C{a_{3D}\over a_{\perp}}\right)^{-1},
\ee
where $C=1.4603\ldots$. In what follows we assume that we are far
from the confinement-induced resonance and the second term in the
brackets in Eq.~(\ref{golsh}) is strictly positive. It is convenient
to introduce a dimensionless parameter $\gamma$, which characterizes
the strength of interaction:
\be
\gamma={mg\over \hbar^2\rho_0},
\ee
where $\rho_0$ is the mean field boson density.

In the absence of tunneling, the Hamiltonian~(\ref{twocond})
corresponds to the Lieb-Liniger model~\cite{LL} of bosons with
point-like interactions. For this model the effective Luttinger
liquid description~\cite{Hal} agrees well with the exact solution
for the ground-state properties~\cite{CCS} and the low-energy
excitations (see Ref.~[\onlinecite{caz}] for the review). We thus
expect that the ``bosonization" procedure underlying the Luttinger
liquid formalism also provides a good description of coupled
condensates, at least when the value of the tunneling $t_{\perp}$ is
not too large. The advantage of this low energy description is that
one can make explicit analytic calculations for both static and
dynamic properties of the system. Physically, the main assumption
justifying use of this effective Hamiltonian is the absence of
strong density fluctuations in the system. At sufficiently short
distances (shorter than the healing length) this condition is
violated. However, as one coarse grains the system, density
fluctuations become weaker and the Luttinger liquid description
becomes justified. Thus the effective formalism can correctly
describe phenomena at wavelengths longer than the healing length of
the condensate.

Within the Luttinger liquid description the Hamiltonian
(\ref{twocond}) splits into parts corresponding to individual
condensates $H_{1,2}$ and the tunneling between them $H_{tun}$. The
first two are characterized by so called Luttinger liquid parameter
$K$ and the sound velocity $v_{s}$, which we assume to be identical
for both condensates:
\beq\label{1,2}
H_{1,2}=\frac{v_{s}}{2}\int dx \left[\frac{\pi}{K}\Pi^{2}_{1,2}(x)+
\frac{K}{\pi}(\partial_{x}\phi_{1,2})^{2}\right].
\eeq
Here the phase field $\phi_{1,2}(x)$ is conjugate to the momentum
$\Pi_{1,2}(x)$.  The relation between the original bosonic operators
 $\Psi^{\dag}_{j}$ and the new fields is given by the bosonization rule
\cite{Hal, caz}
\beq\label{psi-bos}
\Psi^{\dag}_{j}(x)\sim\sqrt{\rho_{0}+
\Pi_{j}(x)}\sum_{m}e^{2im\Theta_{j}(x)}e^{-i\phi_{j}(x)},
\eeq
where $\partial\Theta_{j}(x)/\pi =\rho_{0}+\Pi_{j}(x)$.

In general, the relations of $K$ and $v_s$ to the original
microscopic parameters $m$ and $g$ are not known analytically but
can be deduced numerically from the exact solution of the
Lieb-Liniger model. However, in the limit of large and small
interactions one can obtain approximate expressions~\cite{caz}. Thus
for $\gamma\ll 1$
\beq\label{smallg}
v_s\approx v_{F}{\sqrt{\gamma}\over\pi}\left(1-{\sqrt{\gamma}\over
2\pi}\right)^{1/2},\; K\approx
{\pi\over\sqrt{\gamma}}\left(1-{\sqrt{\gamma}\over
2\pi}\right)^{-1/2}
\eeq
and in the opposite limit $\gamma\gg 1$
\beq\label{largeg}
v_s\approx v_{F}\left(1-{4\over\gamma}\right), && K\approx 1+{4\over
\gamma}.
\eeq
Here $v_{F}=\hbar\pi\rho_0/m$ is the Fermi velocity. Note that
$v_{s}K=v_{F}$ as a consequence of the Galilean invariance. We also
mention that the healing length $\xi_h$ setting the scale of
applicability of Eq.~(\ref{1,2}) is approximately equal to the
inter-particle distance:  $\xi_h\sim 1/\rho_0$ at $\gamma\gg 1$ and
becomes larger: $\xi_h\sim 1/\rho_0\sqrt{\gamma}$ at $\gamma\ll 1$.
We note that experimentally the coupling $\gamma$ can be tuned
through a very wide range~\cite{KWW} by either changing the density
or transverse confinement length or by tuning $a_{3D}$ near the
Feshbach resonance (see Eq.~(\ref{golsh})).

The tunneling part of the Hamiltonian $H_{tun}$ can be described
effectively as
\beq\label{tun}
H_{tun}=-2\Delta \cos(\phi_{1}-\phi_{2}),
\eeq
where we introduced the new parameter $\Delta$. For small
interactions $\gamma\lesssim 1$ we have $\Delta\approx t_\perp
\rho_0$. At strong interactions $\Delta$ will be renormalized and
can significantly deviate from its bare form.

It is convenient to introduce the total and the relative phases
$\phi_{\pm}=(\phi_{1}\pm\phi_{2})/\sqrt{2}$ and
$\Pi_{\pm}=(\Pi_{1}\pm\Pi_{2})/\sqrt{2}$. Then the Hamiltonian
splits into symmetric and antisymmetric parts: $H=H_+ + H_-$, where
\beq
H_{+}&=&{v_s\over 2}\int dx \,\left[{\pi\over
K}(\Pi_+(x))^{2}+{K\over
\pi}(\partial_x\phi_+)^2\right].\nonumber\\
H_{-}&=&{v_s\over 2}\int dx \,\left[{\pi\over
K}(\Pi_-(x))^{2}+{K\over
\pi}(\partial_x\phi_-)^2\right]\nonumber\\
&-&2\Delta\int dx\,\cos(\sqrt{2}\phi_-).
\eeq
Note that the symmetric part of the Hamiltonian $H_+$ in our
approximation is completely decoupled from the antisymmetric part
$H_-$. However, higher nonlinear terms in $\Pi$ and $\partial_x
\phi$ will couple $\phi_+$ and $\phi_-$ so that the symmetric
degrees of freedom can damp excitations of $\phi_-$.

The Hamiltonian $H_-$ can be simplified further. In particular, we
can rescale units of time: $t\to t v_s$ to set $v_s=1$. It is also
convenient to rescale the phase field $\phi$ and the conjugate
momentum: $\phi_-\to \sqrt{\pi/K}\,\phi_-$ and $\Pi_-\to
\sqrt{K/\pi}\,\Pi_-$. Then
\be
H_-=\frac{1}{2}\int dx
[\Pi^{2}_{-}(x)+(\partial_{x}\phi_{-})^{2}-4\Delta\cos(\beta
\phi_{-})].\label{sgham}
\ee
where $\beta=\sqrt{2\pi/K}$. It is also convenient to introduce a
parameter $\xi$:
\be
\xi=\frac{\beta^{2}}{8\pi-\beta^{2}}=\frac{1}{4K-1},
\ee
which we will use later in the text.

\subsection{Quantum SG model: general facts\label{sgfacts}}

The spectrum of the quantum sine-Gordon Hamiltonian (\ref{sgham})
depends on the value of $\beta$. For $4\pi<\beta^{2}<8\pi$ it
consists of solitons and antisolitons. The point $\beta^{2}=4\pi$
corresponds to the free massive fermion theory. In this case
solitons and antisolitons correspond to particles and holes. For
$0<\beta^{2}<4\pi$ the spectrum in addition to solitons and
antisolitons contains their bound states called {\it breathers}.
Note that within the Lieb-Liniger model $K\geq 1$, which implies
that $\beta^2\leq 2\pi$, so that we are always in the latter
regime.~\cite{footnote}. The number of breathers depends on the
interaction parameter $\xi$ and is equal to the integer part of
$1/\xi$. We denote breathers by $B_{n}$ for $n=1\ldots[1/\xi]$. In
the gaussian limit of the sine-Gordon Hamiltonian~(\ref{sgham}),
where one expands $\cos(\beta\phi_-)$ to the quadratic order in
$\phi_-$, there is only one massive excitation corresponding to the
lowest breather $B_{1}$. Solitons (kinks), antisolitons (antikinks)
and breathers are massive particle-like excitations. Soliton and
antisoliton masses in terms of the parameters of the
Hamiltonian~(\ref{sgham}) were computed in
Ref.~[\onlinecite{Zmass}]:
\beq\label{masss}
M_{s}=\left(\frac{\pi\Gamma\left(\frac{1}{1+\xi}\right)\Delta
}{\Gamma\left(\frac{\xi}{1+\xi}\right)}\right)^{(1+\xi)/2}
\frac{2\Gamma\left(\frac{\xi}{2}\right)}{\sqrt{\pi}\,
\Gamma\left(\frac{1+\xi}{2}\right)}.
\eeq
Breather masses are related to the soliton masses via
\beq\label{massb}
M_{B_{n}}=2M_{s}\sin\left(\frac{\pi \xi n}{2}\right).
\eeq
Note that at weak interactions (large $K$ and small $\xi$) the
lowest breather masses are approximately equidistant $M_{B_n}\propto
n$, which suggests that these masses correspond to eigen-energies of
a harmonic theory. There is indeed a direct analogy between the
breathers in the sine-Gordon model and the energy levels of a simple
Josephson junction~\cite{quench}. In the Josephson junction the
energy levels also become approximately equidistant if the
interaction (charging) energy is small.

And finally at $\beta^{2}=8\pi$ the systems undergoes the
Kosterlitz-Thouless transition so that for $\beta^{2}>8\pi$ the
cosine term becomes irrelevant and system is described by the usual
Luttinger liquid.

\subsection{Possible experimental probes}

We consider three different types of modulation experiments, which
can reveal the spectrum of the quantum sine-Grodon model. We
restrict ourselves to the linear response regime, which corresponds
to small perturbations. Modulation of different parameters in the
model focuses on different parts of qSG spectrum and provides
complementary information. A schematic view of possible experimental
setups is given in Fig.~\ref{exp-pic}. The direction of axes of two
parallel condensates is out of the plane of figure.

In this paper we discuss three types of periodic modulations:

a). Modulation of the magnitude of the tunneling amplitude
$\Delta(x,t)$. This kind of modulation couples to the potential
density operator $\cos(\beta\phi)$.

b). Modulation of the relative potential difference between the two
wells $V(x,t)$ . This modulation couples to the momentum operator
$\Pi(x)$, which in turn is proportional to $\partial_{t}\phi$.

c). Modulation of the phase of the tunneling amplitude, e.g.
$\Delta^{*}(x,t)=\Delta\exp(\pm ik(t)x)$. After redefinition of
variables $\phi\rightarrow\tilde{\phi}=\phi\pm k(t)x/\beta$ the term
corresponding to the density of the {\it topological current}
operator $\partial_{x}\phi_{-}$ appears in the Hamiltonian. Note
that the topological charge
\beq\label{topch}
Q=\frac{\beta}{2\pi}\int_{-\infty}^{\infty}\partial_{x}\phi_{-}
\eeq
is conserved and quantized as $\pm n$ with  $n$ being an integer
(corresponding to presence of solitons and antisolitons). We note
that in linear response the modulation of the type c) is equivalent
to modulating the relative current between the two condensates.

For each perturbation of the qSG model that we discussed, one can
define an appropriate correlation function and susceptibility.
According to the fluctuation-dissipation theorem energy absorption
(per unit time) during modulation experiments is given by the
imaginary part of these susceptibilities. The latter are also known
as structure factors. In Sec.~\ref{linrespsec} we present typical
structure factors corresponding to experiments of type a) -
Fig.~\ref{cos1}, type b) - Fig.~\ref{cos2}, and type c) -
Fig.~\ref{dphi2}. A general structure of the response function is
given by a collection of peaks. These peaks can be classified into
two groups: coherent, $\delta$-function-like peaks coming from
single breathers and the broader peaks coming from creating many
(typically two)-particle excitations. From the brief overview given
in Sec.~\ref{sgfacts} it is clear that the absorption spectrum
significantly depends on the interaction parameter $K$. We give a
detailed analysis of structure factors in Sec.~\ref{linrespsec}.

\section{Quantum sine-Gordon model and form-factors of its
operators\label{sgsec}}

The Hilbert space of the sine-Gordon model can be constructed from
the asymptotic scattering states. The latter can be obtained by the
action of operators $A_{a_{k}}(\theta)$ corresponding to elementary
excitations on the vacuum state
\beq\label{asst}
|\theta_{1}\theta_{2}...\theta_{n}\rangle_{a_{1},a_{2},...,a_{n}}=
A^{\dag}_{a_{1}}(\theta_{1})A^{\dag}_{a_{2}}(\theta_{2})...
A^{\dag}_{a_{n}}(\theta_{n})|0\rangle,
\eeq
where the operators $A_{a_{k}}(\theta_{k})$ have internal index
$a_{k}$ corresponding to solitons ($a_{k}=+$), antisolitons
($a_{k}=-$), or breathers ($a_{k}=n$, $n=1\ldots[1/\xi]$) and depend
on the rapidity $\theta_{k}$ which parametrizes the momentum and the
energy of a single-quasi-particle excitation of mass $M_{a_k}$ (see
Eqs.~(\ref{masss}), (\ref{massb})): $E_{a}=M_{a}\cosh(\theta)$,
$P_{a}=M_{a} \sinh(\theta)$. These states are complete and therefore
satisfy the completeness relation:
\beq\label{compl}
1=\sum_{n=0}^{\infty}\sum_{\{a_{i}\} }
\int\prod_{i=1}^{n}\frac{d\theta_{i}}{(2\pi)^{n}n!} |\theta_{n}
\ldots\theta_{1}\rangle_{a_{1}\ldots a_{n}}^{a_{1}\ldots a_{n}}
\langle\theta_{1}\ldots\theta_{n}|.
\eeq
Elementary states or excitations in an integrable field theory can
be described in terms of operators creating or annihilating
asymptotic states. Their commutation relations involve the
scattering matrix $S$
\beq\label{ZFalg}
A_{a}(\theta_{1})A_{b}(\theta_{2}) &=& S^{a'b'}_{ab}
(\theta_{1} - \theta_{2})A_{b'}(\theta_{2})A_{a'}(\theta_{1})\nonumber\\
A_{a}^{\dag}(\theta_{1})A_{b}^{\dag}(\theta_{2}) &=&
S^{a'b'}_{ab}(\theta_{1}-\theta_{2})A_{b'}^{\dag}(\theta_{2}) A_{a'}^{\dag}(\theta_{1})\nonumber\\
A_{a}(\theta_{1})A_{b}^{\dag}(\theta_{2})
&=&2\pi\delta_{ab}\delta(\theta_{1}-\theta_{2})\nonumber\\
&+&S^{b'a}_{ba'}(\theta_{1}-\theta_{2})A_{b'}^{\dag}(\theta_{2})A_{a'}(\theta_{1})
\eeq
These relations are called the Zamolodchikov-Faddeev algebra.  They
generalize canonical commutation relations for bosons and fermions
and reduce to them in some special cases. Since breathers are the
bound states of solitons and antisolitons, the soliton-antisoliton
($a=``+",  b=``-"$) scattering matrix has corresponding imaginary
poles at $\theta_{n}=i\pi(1-n\xi)$, $n=1,\ldots [1/\xi]$.

Different components of $S$-matrix are related to several allowed
scattering processes (related to the configurations in the 6-vertex
model) and as a building block include the following quantity:
\be\label{smat}
S_{0}(\theta)=-\exp\left[-i\int_{0}^{\infty}\frac{dx}{x}
\frac{\sin\left(\frac{2x\theta}{\pi\xi}\right)
\sinh\left(\frac{\xi-1}{\xi}x\right)}{\sinh(x)
\cosh\left(\frac{x}{\xi}\right)}\right].
\ee

\subsection{Form-factors}

The form factors $F^{{\cal O}}$ of a given operator ${\cal O}$ of an
integrable model are the matrix elements in asymptotic states
(\ref{asst}) created by elements $A$ and, $A^\dag$ of the
Zamolodchikov-Faddeev algebra, Eqs. (\ref{ZFalg}). Explicitly,
\beq\label{ffcanon}
F^{{\cal O}}(\theta_{n}\ldots\theta_{1})_{a_{n}\ldots a_{1}}=\langle
0|{\cal O}(0,0)|\theta_{n},\ldots\theta_{1}\rangle_{a_{n}\ldots
a_{1}}.
\eeq
Form-factors (FF) satisfy a set of axioms and functional equations,
which together allow in principle to determine their explicit forms.
Using the so-called crossing relation and the translation invariance
all form-factors can be expressed in terms of matrix elements of the
form (\ref{ffcanon}):
\beq
\langle 0|{\cal
O}(x,t)|\theta_{n},\ldots\theta_{1}\rangle_{a_{n}\ldots
a_{1}}=\exp[i\sum_{j=1}^{n}(E_{j}t+P_{j}x)]\nonumber\\
\times\langle 0|{\cal
O}(0,0)|\theta_{n},\ldots\theta_{1}\rangle_{a_{n}\ldots a_{1}},
\eeq
where $E_{j}=M_{a_{j}}\cosh(\theta_{j})$,
$P_{j}=M_{a_{j}}\sinh(\theta_{j})$ . Note that in the noninteracting
limit of some generic model Zamolodchikov-Faddev algebra reduces
either to Bose or Fermi canonical commutation relations. In this
case form-factors simply reduce to the coefficients of the expansion
of the operator $\mathcal O$ in the second quantized form.

Explicit analytical expressions for the form-factors depend on the
specific type of the operator ${\cal O}$.  In this paper we will be
interested in two particular types of operators: (i) the current
operator in the sine-Gordon model:
\be
J^{\mu}=-\frac{\beta}{2\pi}\epsilon^{\mu\nu}\partial_{\nu}\phi_{-},
\ee
where $\mu=0,1$ correspond to time $t$ and space $x$ components; and
(ii) the other operator corresponding to the trace of the
stress-energy tensor
$T^{\mu\nu}=\partial_{\mu}\partial_{\nu}\phi-\delta_{\mu\nu}{\cal
L}$, where ${\cal L}$ is the Lagrangian,
\beq
T\equiv Tr(T^{\mu}_{\nu})&=&4\Delta\cos(\beta\phi_{-}).
\eeq
These two operators have different properties with respect to the
Lorentz group and different charge parity. Therefore their
properties can be distinguished by the topological charge
(\ref{topch}) and the charge conjugation operator $C$ defined by
\beq
C|0\rangle=|0\rangle,\quad
CA_{s}^{\dag}(\theta)C^{-1}&=&A_{\bar{s}}^{\dag}(\theta),\\
CA_{B_{n}}^{\dag}(\theta)C^{-1}&=&(-1)^{n}A_{B_{n}}^{\dag}(\theta)
\eeq
The action of both the current operator and $\cos(\beta\phi_{-})$ do
not change the value of $Q$.

There is extensive literature on computation of form-factors for the
quantum sine-Gordon model. To our knowledge such analysis was
initiated in Ref.~[\onlinecite{VG}] and extended later in
Ref.~[\onlinecite{W}]. These developments are summarized in the
book~[\onlinecite{S}] and  more recently in Refs.~[\onlinecite{L}],
[\onlinecite{Lu}], [\onlinecite{BK}].

If there are bound states in the theory, like breathers in our case,
the form-factors which include these bound states can be constructed
using the residue at the poles of the soliton-antisoliton
form-factors. To get form factors corresponding to higher breathers
one can use a so-called fusion procedure. The receipt is that the
breather $B_{n+m}$ appears as a bound state of the $B_{n}$ and
$B_{m}$ breathers, $B_{n}+B_{m}\rightarrow B_{n+m}$. Thus the
residue of the pole in the form-factor $F_{B_{n}B_{m}}$ will
correspond to the form-factor $F_{B_{n+m}}$. For example, to
construct the form factor which includes the second breather $B_{2}$
we can use the form factor with $n+2$ breathers $B_{1}$ and the
fusion formula
\beq
i\mbox{Res}_{\epsilon\rightarrow 0}F_{a_{1}\ldots
a_{n}11}\biggl(\theta_{1},\ldots,\theta_{n},\theta_{n+1}+iU_{12}^{1}
-\frac{\epsilon}{2},\nonumber\\\theta_{n+1}-iU_{12}^{1}
+\frac{\epsilon}{2}\biggr) =\Gamma^{2}_{11}F_{a_{1}\ldots
a_{n}2}(\theta_{1},\ldots,\theta_{n},\theta_{n+1}),
\eeq
where the fusion angle $U^{1}_{12}=\pi\xi/2$ and the 3-particle
coupling $\Gamma^{2}_{11}=\sqrt{2\tan(\pi\xi)}$ is given by the
residue of the S-matrix (\ref{smat}).

\subsubsection{Soliton-antisoliton form-factors}
1). ${\cal O}=e^{i\beta\phi}$. For the sine-Gordon model different
form factors for the phase operators $e^{i\beta\phi}$ have been
found in Refs.~[\onlinecite{S,L}]. In particular, the nonvanishing
$``+-"$ (soliton-antisoliton) form factor is given by
\beq\label{sscos}
&& \langle 0|e^{i\beta\phi(0,0)}|A^{\dag}_{\pm}(\theta_{2})
A^{\dag}_{\mp}(\theta_{1})\rangle={\cal G}_{\beta}
\exp[I(\theta_{12})]\\
&&~~~~~\times\frac{2i\cot\left(\frac{\pi\xi}{2}\right)
\sinh\left(\theta_{12}\right)}
{\xi\sinh\left(\frac{\theta_{12}+i\pi}{\xi}\right)}
\exp\left({\mp\frac{\theta_{12}+i\pi}{2\xi}}\right),\nonumber
\eeq
where $\theta_{12}=\theta_{1}-\theta_{2}$, and
\beq\label{I}
I(\theta)=\int_{0}^{\infty}\frac{dt}{t}\frac{\sinh^{2}
[t(1-\frac{i\theta}{\pi})]\sinh[t(\xi-1)]}{\sinh[2t]\cosh[t]\sinh[t\xi]}
\eeq
Equations~(\ref{sscos}) as well as the normalization of general
$\cos(\beta\phi_{-})$ -type form-factors includes the following
vacuum-vacuum amplitude:~\cite{LZ}
\beq\label{vacev}
&&{\cal G}_{\beta}=\langle0|e^{i\beta\phi}|0\rangle\nonumber\\
&&~~~~=\Delta^\xi
\frac{(1+\xi)\tan\left(\frac{\pi\xi}{2}\right)\Gamma^{2}\left(\frac{\xi}{2}\right)}
{2\pi\,\Gamma^{2}\left(\frac{1+\xi}{2}\right)}
\left[\frac{\pi\Gamma\left(\frac{1}{1+\xi}\right)}
{\Gamma\left(\frac{\xi}{1+\xi}\right)}\right]^{1+\xi}\!\!\!.\phantom{XX}
\eeq

Higher order soliton-antisoliton form-factors (e.g. with two
solitons and two antisolitons) are also available in the
literature~\cite{S}. For generic values of $K$ they are given by
multiple integrals, whereas for the half-integer values of $4K$ (in
our notations) the expressions considerably simplify. In the next
section we explain that the relative contribution to the operator's
expectation value from the states which include many
soliton-antisoliton pairs is very small and can be neglected.

2). ${\cal O}=\partial_{\mu}\phi_{-}$. The form-factors for these
operators can be deduced from the known form-factors of the current
operator $J^{\mu}$. In particular, the form-factor for the operator
$\partial_{t}\phi_{-}$ is
\beq\label{curkak}
&&F_{+,-}^{\partial_{t}\phi_{-}}(\theta_{1},\theta_{2})
=-F_{-,+}^{\partial_{t}\phi_{-}}(\theta_{1},\theta_{2})\\
&&=\frac{4\pi^{3/2}\sqrt{2}
M_{s}}{\beta}\frac{\cosh\left(\frac{\theta_{1}+\theta_{2}}{2}\right)
\sinh\left(\frac{\theta_{1}-\theta_{2}}{2}\right)}
{\cosh\left(\frac{\theta_{1}-\theta_{2}+i\pi}{2\xi}\right)}I(\theta_{12}).\nonumber
\eeq
The FF for the  $\partial_{x}\phi_{-}$ is given by
\beq\label{curkakx}
&&F_{+,-}^{\partial_{t}\phi_{-}}(\theta_{1},\theta_{2})
=-F_{-,+}^{\partial_{t}\phi_{-}}(\theta_{1},\theta_{2})\\
&&=\frac{4\pi^{3/2}\sqrt{2}
M_{s}}{\beta}\frac{\sinh\left(\frac{\theta_{1}+\theta_{2}}{2}\right)
\sinh\left(\frac{\theta_{1}-\theta_{2}}{2}\right)}
{\cosh\left(\frac{\theta_{1}-\theta_{2}+i\pi}{2\xi}\right)}I(\theta_{12}).\nonumber
\eeq
Multi soliton-antisoliton form factors of this operator can be found
in Ref.~[\onlinecite{S}] in terms of integrals. Their relative
contribution is small as well~\cite{CET} and we will neglect them.

\subsubsection{Breather form-factors}

1). ${\cal O}=e^{i\beta\phi}$. The one-breather form factors of the
type $\langle 0|e^{i\beta\phi}|B_{n}(0)\rangle\equiv F_{B_{n}}$ can
be computed from the residue of the soliton-antisoliton form-factors
at points $\theta_{n}=i\pi(1-n\xi))$, or, equivalently, from the
procedure of fusion of several breathers. For example the breather
$B_{2}$ is the bound state of two breathers $B_{1}$. This procedure
is known as a bootstrap approach.
\be\label{1B}
F_{B_{n}}^{\exp(i\beta\phi)}\!\!=\!\frac{{\cal
G}_{\beta}\sqrt{2}\cot(\frac{\pi\xi}{2})\sin(\pi
n\xi)\exp[I(-\theta_{n})]e^{\frac{i\pi
n}{2}}}{\sqrt{\cot(\frac{\pi\xi
n}{2})\prod_{s=1}^{n-1}\cot^{2}(\frac{\pi \xi s}{2})}}.
\ee
We note that Eq.(\ref{1B}) reveals the parity property of breathers:
odd breathers are antisymmetric with respect to the charge symmetry
transformation, whereas even breathers are symmetric.

The form-factors of several breathers can be derived as well. Since
the breather $B_{1}$ is like a fundamental particle in a theory, all
higher-level breathers form-factors can be expressed via a fusion
procedure of $B_{1}$-form-factors. Because of the parity the
non-zero form-factors for the operator $\cos(\beta\phi)$ contain
either even breathers or even number of odd breathers. It is thus
important to have an explicit expression for the $B_{1}\ldots
B_{1}$-form factors:
\beq\label{1B-n}
F_{B_{1}\ldots
B_{1}}^{cos(\beta\phi)}=\langle0|\cos(\beta\phi)|B_{1}(\theta_{2n})\ldots
B_{1}(\theta_{1})\rangle \nonumber\\
=\frac{1}{2}{\cal G}_{\beta}\lambda^{2n}\prod_{1\leq i<j\leq
2n}{\cal
R}(\theta_{k}-\theta_{j})\frac{\det(\Sigma_{ij}')}{\det(\Sigma_{ij})}.
\eeq
Here the $(2n-1)\times(2n-1)$ matrices
\begin{displaymath}
\Sigma_{ij}=\sigma_{2i-j},  \quad
\Sigma_{ij}'=\sigma_{2i-j}{\sin[\pi\xi(i-j+1)]\over \pi\xi}
\end{displaymath}
are expressed in terms of symmetric polynomials
\begin{displaymath}
\sigma_{k}\equiv\sigma_{k}^{(2n-1)}=
\sum_{i_{1}<i_{2}<\ldots<i_{k}}^{2n-1}x_{i_{1}}\ldots x_{i_{k}}
\end{displaymath}
with $x_{i}=e^{\theta_{i}}$,
\beq\label{J-R}
\lambda&=&2\cos\left[\frac{\pi\xi}{2}\right]\sqrt{2\sin\left[\frac{\pi\xi}{2}\right]}
\exp\left[-\int_{0}^{\pi\xi}\frac{dt}{2\pi}\frac{t}{\sin t}\right],\nonumber\\
{\cal R}(\theta)&=&N\exp\biggl[8\int_{0}^{\infty}\frac{dt}{t} \frac{\sinh(t)\sinh(t\xi)\sinh(t(1+\xi))}{\sinh^{2}(2t)}\nonumber\\
&&~~~~~~~~~~~~~~~~~~~~~~~~\times \sinh^{2}\left(1-i\frac{\theta}{\pi}\right)\biggr],\\
N&=&\exp\left[4\int_{0}^{\infty}\frac{dt}{t}\frac{\sinh(t)\sinh(t\xi)\sinh(t(1+\xi))}{\sinh^{2}(2t)}\right].\nonumber
\eeq
Here the function ${\cal R}(\theta)$ satisfies a useful relation
which allows to evaluate its residue at $\theta=-i\pi(1-n\xi)$
\beq
{\cal R}(\theta){\cal R}(\theta\pm
i\pi)=\frac{\sinh(\theta)}{\sinh(\theta)\mp i\sinh(\pi\xi)}.
\eeq
 The formula
(\ref{1B-n}) can be shown to agree both with the sine-Gordon
``bosonization" by Lukyanov~\cite{Lu} and the Bethe-Ansatz-based
method of Ref.~[\onlinecite{BK}].

The form-factor for the $B_{2}-B_{2}$ breathers can be derived from
Eq.~(\ref{1B-n}) using the fusion procedure:
\beq\label{b2b2cos}
&&F_{B_{2}B_{2}}^{\cos(\phi)}(\theta)=\frac{{\cal
G}_{\beta}\lambda^{4}}{2}\tan(\pi\xi)
\left(1+\frac{1}{\cosh(\theta)+\cos(\pi\xi)}\right)\nonumber\\
&&~~~~~~~~~~~~~~\times\frac{{\cal R}^{2}(\theta){\cal
R}(-\theta-i\pi\xi){\cal R}(-\theta+i\pi\xi)}{{\cal
R}(-i\pi(1+\xi))}\,.\nonumber
\eeq

2). ${\cal O}=\partial_{t}\phi_{-}$. Because of the charge
reflection symmetry, odd-type one-breather form-factors exist for
this operator. They can be found either directly from the poles of
the soliton-antisoliton form-factor (\ref{curkak}) or via the
relation $\langle 0|\partial_{t}\phi_{-}|B_{n}(\theta)\rangle
=\lim_{a\rightarrow 0}\frac{1}{ia}M_{B_{n}}\cosh(\theta)\langle
0|e^{ia\phi_{-}}|B_{n}(\theta)\rangle$. The result is
\beq\label{1b-J}
F_{B_{m}}^{\partial_{t}\phi}(\theta)&=&\left[\cot\left(\frac{\pi
m\xi}{2}\right)\prod_{s=1}^{m-1}\tan^{2}\left(\frac{\pi s\xi}{2}\right)\right]^{1/2}\\
&\times&\frac{M_{B_{m}}\sqrt{2}(2\pi)^{3/2}\xi}{\beta}\exp(I(-\theta_{m}))\cosh(\theta),\nonumber\\
F_{B_{m}}^{\partial_{x}\phi}(\theta)&=&\left[ \cot\left(\frac{\pi
m\xi}{2}\right) \prod_{s=1}^{m-1}\tan^{2} \left(\frac{\pi s\xi}{2}\right)\right]^{1/2}\\
&\times
&\frac{M_{B_{m}}\sqrt{2}(2\pi)^{3/2}\xi}{\beta}\exp(I(-\theta_{m}))\sinh(\theta),\nonumber
\eeq
where $m$ is odd.

To compute form-factors for several breathers we note that there is
a correspondence between the lowest $B_{1}$ breather sector of the
sine-Gordon theory and the {\it sinh}-Gordon theory. Using this
correspondence the $B_{1}$ - form-factors can be obtained via the
recursion formula~\cite{FMS} from the form-factors of
$\varphi_{\sinh}$ - operator in the {\it sinh}-Gordon model. This
recursion gives us odd-number non-zero $B_{1}$ form-factors
\beq
&&F^{\partial_{t}\phi_{-}}_{B_{1}^{(1)}\ldots B_{1}^{(2n+1)}}=
\frac{(2\pi)^{3/2}\lambda^{2n+1}\xi}{\beta\sin(\pi\xi)}M_{B_{1}}
\left[\sum_{n=1}^{2n+1}\cosh(\theta_{n})\right]\nonumber\\
&&\times\sigma_{2n+1}^{(2n+1)}P_{2n+1}(x_{1},\ldots
x_{2n+1})\prod_{i<j}\frac{{\cal
R}(\theta_{i}-\theta_{j})}{x_{i}+x_{j}},
\eeq
where $x_{i}=e^{\theta_{i}}$ and first polynomials $P$ are given
by\cite{FMS}
\beq
P_{3}(x_{1},x_{2},x_{3})&=&1\\
P_{5}(x_{1},\ldots,x_{5})&=&\sigma_{2}\sigma_{3}-4\cos^{2}(\pi\xi)\sigma_{5}
\eeq
These formulas allow us to compute, for example, the $B_{1}-B_{2}$
form-factor, which is the first nontrivial 2-particle form-factor
after the soliton-antisoliton form-factor:
\beq\label{b1b2cur}
&&F_{B_{1}B_{2}}^{\partial_{x}\phi}(\theta_{1},\theta_{2})=(M_{B_{1}}\sinh(\theta_{1})
+M_{B_{2}}\sinh(\theta_{2}))\nonumber\\
&&~~~~~~~~~~~~~~~~~\times\tilde{F}_{B_{1}B_{2}}^{\partial_{x}\phi}(\theta_{12}),\nonumber\\
&&\tilde{F}_{B_{1}B_{2}}^{\partial_{x}\phi}(\theta_{12})={\pi^{3/2}\lambda^{3}\xi\over
\sqrt{8}\sin(\pi\xi)\beta}\nonumber\\
&&\times\frac{\sqrt{\tan(\pi\xi)}\,{\cal
R}(\theta_{23}+\frac{i\pi\xi}{2}){\cal
R}(\theta_{23}-\frac{i\pi\xi}{2})}
{\cos(\frac{\pi\xi}{2})\left(\cosh(\theta)+\cos(\frac{\pi\xi}{2})\right){\cal
R}\left(-i\pi(1+\xi)\right)}.\nonumber
\eeq
It is remarkable that the further fusion of $B_{1}$ and $B_{2}$
particles produces the $B_{3}$ form-factor, which is identical to
the result from Eq. (\ref{1b-J}). This agreement is a direct
manifestation of the {\it bootstrap}~\cite{ZZ}.

\section{Linear response theory for coupled condensates\label{linrespsec}}

Experimentally one can modulate the relative coupling between two
condensates by applying small modulations in space and/or time.
Moreover one can change the densities in both condensates such that
the relative density can vary. Following the previous discussion we
will consider three relevant operators which describe different
physical perturbations
\beq
&a).& {\cal O}_{T}(x,t)=\cos(\beta\phi_{-}),\\
&b).& {\cal O}_{J_{1}}(x,t)=\partial_{t}\phi_{-},\\
&c).& {\cal O}_{J_{0}}(x,t)=\partial_{x}\phi_{-}
\eeq
These operators directly correspond to the experimental setups (a),
(b) and (c) discussed in the Introduction. The conserved topological
charge in these notations is given by the integral of $J^{0}$.

In the linear response regime we are interested in computing
two-point correlation functions
\beq
\langle {\cal O}(x,t){\cal O}^{\dag}(0,0)\rangle,
\eeq
where the expectation value is taken over the ground state of the
unperturbed system. Inserting the completeness relation
(\ref{compl}) and using the relativistic invariance allows us to
express this average in the following weighted sum over the
intermediate states:
\beq
&&\langle {\cal O}(x,t){\cal O}^{\dag}(0,0)\rangle
=\sum_{n=0}^{\infty}\sum_{\{a_{i}\} } \int\prod_{i=1}^{n}
\frac{d\theta_{i}}{(2\pi)^{n}n!}\nonumber\\
&&\times\mathrm
e^{i\sum_{j=1}^{n}P_{j}x-E_{j}t}\,\left|\langle0|{\cal
O}(0,0)|\theta_{n}\ldots\theta_{1}\rangle_{a_{n}\ldots
a_{1}}\right|^{2}.
\eeq
It is convenient to define the dynamical structure factor as  a
Fourier transform of this quantity,
\beq\label{sqw}
&&S^{{\cal O}}(q,\omega)=Im[\int\int_{-\infty}^{\infty}dxdt
e^{i\omega t-iqx}\langle {\cal O}(x,t){\cal
O}^{\dag}(0,0)\rangle]\nonumber\\
&&=2\pi\sum_{n=0}^{\infty}\sum_{a_{i}}\int\prod_{i=1}^{n}\frac{d\theta_{i}}{(2\pi)^{n}n!}|F^{{\cal
O}}(\theta_{1}\ldots\theta_{n})_{a_{1}\ldots a_{n}}|^{2}\nonumber\\
&&\times\delta\biggl(q-\!\sum_{j}M_{a_{j}}\sinh\theta_{j}\biggr)
\delta\biggl(\omega-\!\sum_{j}M_{a_{j}}\cosh\theta_{j}\biggr),
\eeq
where $F^{{\cal O}}$ is the corresponding form-factor of the
operator ${\cal O}$. We comment here on the structure of the
expression (\ref{sqw}). First, it is represented as a sum of
contributions coming from different excited states of the
Hamiltonian. Second, it is known from rather general phase-space
arguments~\cite{Muss, DMS} that the sum over intermediate states of
the form (\ref{sqw}) converges rapidly. In particular, it was shown
in Ref.~[\onlinecite{CET}] that the two-soliton-two-antisoliton
contribution is somewhat 500 smaller than the soliton-antisoliton
one. Therefore, in general, the contributions with small number of
excited particles will dominate the result . The expression for the
structure factor (\ref{sqw}) can be also rewritten as the following
sum
\beq
S^{{\cal O}}(q,\omega)& =& S^{{\cal
O}}_{(1)}(q,\omega)+\frac{1}{4\pi}S^{{\cal
O}}_{(2)}(q,\omega)+\frac{1}{24\pi^{2}}S^{{\cal
O}}_{(3)}(q,\omega)\nonumber\\
&+&\ldots
\eeq
The specific form of $S^{{\cal O}}_{(n)}(q,\omega)$ significantly
depends on the charge parity of the operator and the topological
charge parity. The current operator is odd with respect to the
$C$-parity whereas $\cos(\beta\phi_{-})$ is even. All operators we
consider do not change the topological charge.

Explicit expressions for the $S^{\cos(\beta\phi)}(q,\omega)$ have
already appeared in literature \cite{ET,EK} in the context of 1D
spin-1/2 systems discussed in the Introduction. We present them here
for completeness.

\begin{figure}[h]
\begin{center}
\includegraphics[width=9cm,angle=0]{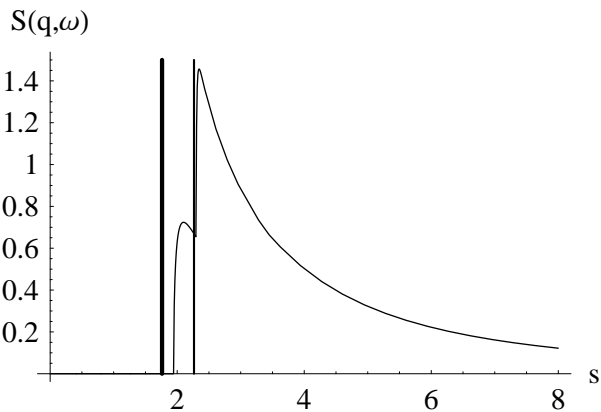}
\caption{ Structure factor for the experiment of type (a)
corresponding to the phase operator $\cos(\beta\phi)$ for the
Luttinger parameter $K=1.15$ and the tunneling gap $\Delta=0.1$.
Peaks correspond to the following leading contributions (from left
to right): a single breather $B_{2}$; breathers $B_{1}B_{1}$; and
soliton-antisoliton $A_{+}A_{-}$. }
\label{cos1}
\end{center}
\end{figure}
\begin{figure}[t]
\begin{center}
\includegraphics[width=8.5cm,angle=0]{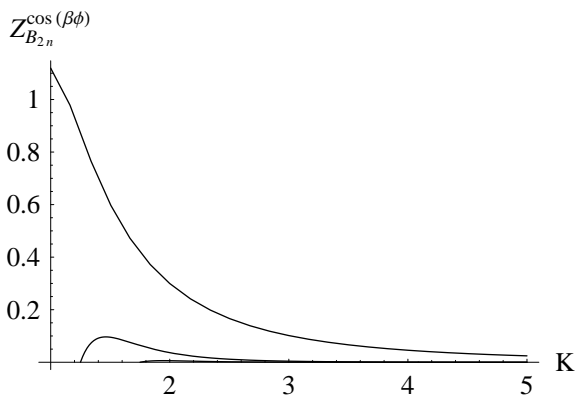}
\caption{ Spectral weights $|F_{B_{n}}^{\cos(\phi)}|^{2}$ of
single-breather contributions corresponding to the scheme a). Shown
here are contributions for (from top to bottom) $B_{2}$, $B_{4}$ and
$B_{6}$ (very small). }
\label{spw-cos}
\end{center}
\end{figure}


a). In this case ${\cal O}=\cos(\beta\phi_{-})$. Then the single
particle contribution to the structure factor is given by
\beq
S^{\cos(\beta\phi)}_{(1)}(q,\omega)=\sum_{n=1}^{[1/\xi]}{\cal
Z}_{B_{2n}}^{\cos(\beta\phi)}\delta(s^{2}-M_{B_{2n}}^{2}),
\eeq
where $s^2=\omega^2-q^2$. Thus $S_{(1)}$ corresponds to excitation
of isolated even breathers. Here the spectral weights are
\beq
{\cal
Z}_{B_{2n}}^{\cos(\beta\phi)}=|F_{B_{2n}}^{\cos(\beta\phi)}|^{2},
\eeq
where $F_{B_{2n}}^{\cos(\beta\phi)}$ is given in Eq.~(\ref{1B}). So
this contribution corresponds to the absorbtion peaks at
$\omega=\omega_{B_{2n}}$ (for $q=0$). The spectral weights are
illustrated in Fig.~(\ref{spw-cos}) for $n=1,2,3$. Note that these
weights decrease with increasing $n$ as well as with increasing $K$.
This plot suggests that breathers with small $n$ dominate the
absorbtion.

\begin{figure}[h]
\begin{center}
\includegraphics[width=9cm,angle=0]{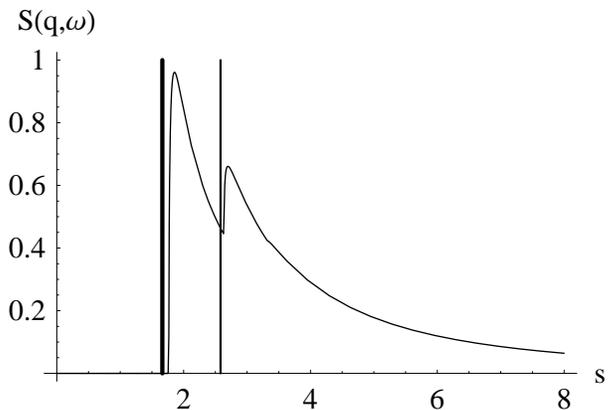}
\caption{ Structure factor for the experiment of type a)
corresponding to the phase operator $\cos(\beta\phi)$ for $K=1.4$
and tunneling $\Delta=0.1$. Peaks correspond to the following
leading contributions (from left to right): single breather $B_{2}$;
breathers $B_{1}B_{1}$; single breather $B_{4}$; and
soliton-antisoliton $A_{+}A_{-}$. }
\label{cos2}
\end{center}
\end{figure}

The two particle contribution to the structure factor corresponding
to excitation of particles $A_{1}$ and $A_{2}$ with masses
$M_{A_{1}}$ and $M_{A_{2}}$ can be generally expressed as
\be
S^{{\cal
O}}_{(2)}(q,\omega)=Re\left[\frac{|F_{A_{1}A_{2}}(\theta_{12})|^{2}}
{\sqrt{\left(s^{2}-M_{A_{1}}^{2}-M_{A_{2}}^{2}\right)^{2}
-4M_{A_{1}}^{2}M_{A_{2}}^{2}}}\right],
\ee
where
\beq\label{12}
\theta_{12}=\theta(q,\omega)=\mbox{arccosh}
\left(\frac{s^{2}-M_{A_{1}}^{2}-M_{A_{2}}^{2}}{2M_{A_{1}}M_{A_{2}}}\right).
\eeq
Direct substitution of the single breather form-factor, Eq.
(\ref{1B}), the soliton-antisoliton form-factor (Eq.(\ref{sscos})),
and breather-breather form-factors (Eqs.
(\ref{1B-n}),(\ref{b2b2cos})) from the previous section gives the
results for the corresponding structure factors. Note that the
soliton-antisoliton contribution has to be weighted by a factor of
2. The results are shown in Figs.~\ref{cos1} and \ref{cos2} for some
fixed values of the Luttinger parameter $K$ and the inter-chain
coupling $\Delta$. We see that the generic form of the structure
factor (and this of the absorbtion peaks) has a coherent part
represented by a sequence of $\delta$-peaks corresponding to
coherent isolated even breather contributions and the incoherent
part corresponding to different two-particle contributions. In
Figs.~\ref{cos1} and \ref{cos2} we show the contributions from
$B_{1}B_{1}$, $A_{+}A_{-}$, $B_{2}B_{2}$ (incoherent part) and
$B_{2}$, $B_{4}$ (coherent peaks) form-factors.

\begin{figure}[t]
\begin{center}
\includegraphics[width=8.5cm,angle=0]{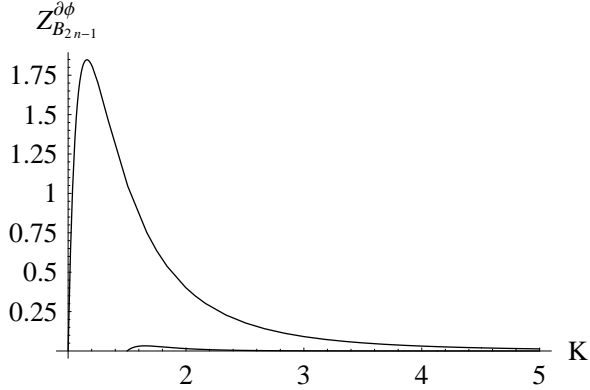}
\caption{ Spectral weights
$|F_{B_{n}}^{\partial_{t}\phi)}|/E_{B_{n}}$
($|F_{B_{n}}^{\partial_{t}\phi)}|/P_{B_{n}}$)of single-breather
contributions appearing in setups b) and c). Shown here are
contributions for (from top to bottom) $B_{1}$, $B_{3}$ and $B_{5}$.
}
\label{dphi1}
\end{center}
\end{figure}

b). For ${\cal O}=\partial_{t}\phi_{-}$ the general form of the
structure factor is different:
\be\label{curz}
S^{\partial_{t}\phi}_{(1)}(q,\omega)={\cal
Z}_{B_{2n-1}}^{\partial_{t}\phi}\omega^{2}
\delta\left(s^{2}-M_{B_{2n-1}}^{2}\right),
\ee
where ${\cal
Z}_{B_{2n-1}}^{\partial_{t}\phi}=|F_{B_{2n-1}}^{\partial_{t}\phi}/E_{B_{2n-1}}|^{2}$
is given by Eq. (\ref{1b-J}). The spectral weights ${\cal
Z}_{B_{2n-1}}^{\partial_{t}\phi}$ are compared in Fig.~\ref{dphi1}
for $n=1,2,3$.

\begin{figure}[t]
\begin{center}
\includegraphics[width=8.5cm,angle=0]{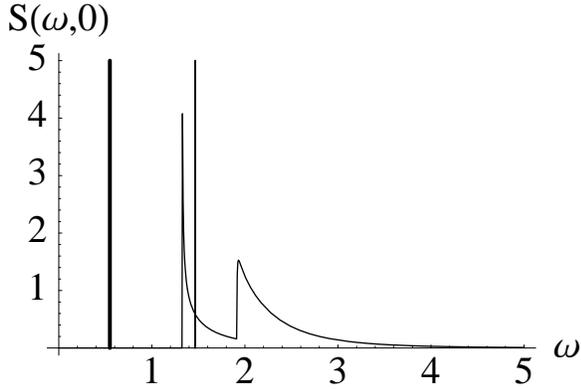}
\caption{ Structure factor $S(w,q=0)$ for the setup scheme of type
(b) corresponding to operator ${\cal O}=\partial_{t}\phi$ at $K=1.6$
and $\Delta=0.05$. The $\delta$-peaks (solid bold line)
corresponding to the breathers $B_{1}$ (thick line) and $B_{3}$
(thin line) are coherent contributions whereas peaks in the
incoherent background (from left to right) correspond to the
breather $B_{1}-B_{2}$ and soliton-antisoliton contributions.}
\label{dphi2}
\end{center}
\end{figure}

\begin{figure}[t]
\begin{center}
\includegraphics[width=8.5cm,angle=0]{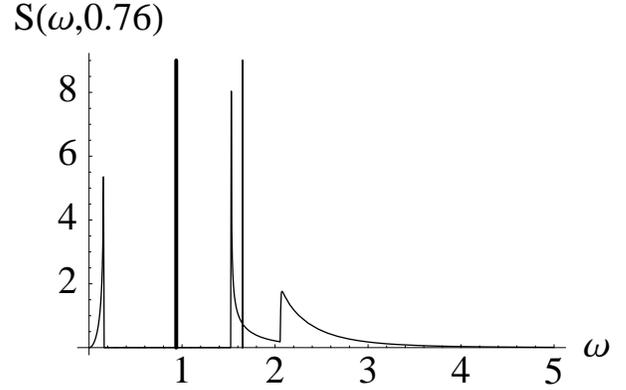}
\caption{ Structure factor $S(w,q=0.76)$ for the setup scheme of
type (b) corresponding to operator ${\cal O}=\partial_{t}\phi$ at
$K=1.6$ and $\Delta=0.05$. The structure of peaks is the same as in
Fig.~\ref{dphi2} except for the appearance of the satellite
breathers peak at small $\omega$. It starts from nonzero $q$ and
exists further for all $q$'s. }
\label{dphi3}
\end{center}
\end{figure}

Similarly the soliton-antisoliton contributions are
\be\label{curss}
S^{\partial_{t}\phi}_{(2),s\bar{s}}(q,\omega)=
Re\left[\frac{\omega^{2}\sqrt{s^{2}-4M_{s}^{2}}}{s^{3}}\frac{|I(\theta_{12})|^{2}}
{\cosh\left(\frac{\theta_{12}}{\xi}\right)+\cos\left(\frac{\pi}{\xi}\right)}\right],
\ee
where $\theta_{12}$ is defined in Eq.~(\ref{12}) with
$M_{A_{1}}=M_{A_{2}}=M_{s}$ and $I(\theta)$ is given by Eq.
(\ref{I}). And finally the two breather contributions are
\beq\label{cur12}
&&S^{\partial_{t}\phi}_{(2),B_{1}B_{2}}(q,\omega)=\nonumber\\
&&Re\left[\frac{\omega^{2}|\tilde{F}(\theta_{12})|^{2}}
{\sqrt{(s^{2}-M_{B_{1}}^{2}-M_{B_{2}}^{2})^{2}-(2M_{B_{1}}M_{B_{2}})^{2}}}\right],
\eeq
where $\theta_{12}$ is defined in (\ref{12}) and
$\tilde{F}(\theta_{12})$ is introduced in (\ref{b1b2cur}). In
Fig.~\ref{dphi2} we show the structure factor for the operator
${\cal O}=\partial_{t}\phi_{-}$ for $K=1.6$. As before we include
contributions corresponding to excitations of $B_{1}$, $B_{3}$,
$A_+A_-$, and $B_{1}B_{2}$.

c). The structure factor for this type of modulation is very much
the same as for the case b). In particular, the $\omega$-dependence
in the nominator of Eqs.~(\ref{curz}-\ref{cur12}) should be replaced
by $q$.

\section{Summary and Outlook}
\label{sumsec}

We considered a system of two coupled one dimensional condensates.
We showed that the low-energy dynamics of this system can be
described by the quantum sine-Gordon model (Sec.~\ref{effsec}). This
model is integrable and supports collective excitations (solitons,
anti-solitons and breathers). To reveal these excitations we propose
to study modulations of the tunneling amplitude - type a), of the
population imbalance - type b) and of the tunneling phase - type c).
Corresponding experiments provide complementary information about
the structure of excitations of the quantum sine-Gordon model. The
modulations of type a) reveal {\it even}-number coherent breather
peaks and even two-breather contributions whereas modulations of
type b) and type c) couple to the odd sector of the spectrum, so
that the coherent peaks correspond to {\it odd} breathers and odd
two-particle excitations. All types of modulations show the
soliton-antisoliton response. The effect of the Luttinger
interaction parameter $K$ of the individual condensates is two-fold:
(i) the spectrum content is entirely determined by the strength of
$K$, as it is explained in Sec.~\ref{sgfacts}; and (ii) the spectral
weights of coherent single-particle excitations $B_n$ with $n\geq 2$
decrease significantly with increasing $K$. This result is not
surprising since at large $K$ we anticipate that the lowest energy
excitations are well described by the Gaussian model of the massive
scalar field. In this limit there are only massive phonon-like
excitations corresponding to $B_1$. For the type a) modulation these
excitations can be created only in pairs giving raise to the
continuum contribution to the spectral function. While for type b)
and type c) modulations the lowest energy contribution comes from
excitations of isolated $B_1$ breathers and then there is a
three-particle continuum. We note that in the weakly interacting
limit $K\gg 1$ the isolated $B_1$ peak corresponds to exciting
Josephson oscillations between the two condensates. Also we note
that the soliton-antisoliton contribution to the spectral function
rapidly diminishes with increasing $K$. . The experimental
visibility of various contribution will also be determined by the
tunneling strength $\Delta$. In general one can observe an
approximate scaling of structure factors with increasing $\Delta$.

We comment that the idea behind modulation experiments is similar to
exciting parametric resonance in a usual harmonic oscillator. This
analogy becomes even more transparent for the special case of zero
dimensional condensates, which is equivalent to the Josepphson
junction. At weak nonlinearity the Josephson junction in turn is
equivalent to a harmonic oscillator. The type a) modulation of the
tunneling amplitude is analogous to the modulation of the mass of
this oscillator and the type b) modulation is similar to the
modulation of the equilibrium position of this oscillator. The
strongest parametric excitation of a harmonic oscillator with
frequency $\omega_0$ occurs at $\omega=2\omega_0$ for the modulation
of type a) and at $\omega=\omega_0$ for the modulation of the type
b).  These transitions, in turn, correspond to excitations of the
second -a) and the first - b) energy levels in the Josephson
junction. As we discussed in Ref.~[\onlinecite{quench}] the first
energy level of the Josephson junction corresponds to the $B_1$
breather in the sine-Gordon model, the second energy level does to
the $B_2$ breather and so on. And indeed as we showed above the
contributions from isolated $B_2$ and $B_1$ breathers to the
absorbtion are dominant for the type a) and type b) modulations
respectively. We can also come to similar conclusions using the
classical analysis\cite{quasicl,Chirikov} of the perturbed SG model
for small $\xi$ limit. One needs to bear in mind that here we deal
with a nonlinear system and the parametric resonance strictly
applies to a harmonic oscillator. So the response of the system to
modulations is more complicated. However, qualitatively the picture
remains very similar in this limit.

Experimentally the absorbtion can be enhanced by increasing the
magnitude of the modulation signal. Even though our analysis is
limited only to weak perturbations, where one can use the linear
response, we expect that the overall picture will not significantly
change in the nonlinear regime. However, one always needs to make
sure that the nonlinear effects do not cause various instabilities
in the system. For example, the modulation of type c) is limited by
the possible commensurate-incommensurate phase transition~\cite{JN},
which occurs if the breather's gap gets comparable with the change
of the chemical potential per particle. Another effect which exists
on the classical level is a dissociation of breathers into decoupled
soliton-antisoliton pair for relatively strong
perturbations\cite{KMS}.

It is important to realize that since we deal with an integrable or
nearly integrable system the energy absorbtion is not necessarily
related to the loss of the phase coherence, which is usually
measured in standard time of flight experiments. Indeed, the two
quantities are related if the absorbed energy is quickly
redistributed among all possible degrees of freedom. However, in
integrable or nearly integrable model this is not the case. For
example, in a recent experiment by Kinoshita {\em et.
al.}~\cite{KWW2} it was demonstrated that there is no thermalization
in a single one-dimensional Bose gas even after extremely long
waiting times. Thus one possibility to measure the absorbtion is to
slowly drive the excited system into the non-integrable regime where
the thermalization occurs and then perform conventional
measurements. The other possibility is to directly measure the
susceptibilities. For example, for the type a) modulation one can
measure the expectation value of the phase difference between the
condensates. For the type b) modulation one can measure the
population imbalance between the two condensates and for the type c)
modulation one can measure the relative current between the two
condensates. In principle, one can always measure the loss of the
phase contrast as it was done in Ref.~[\onlinecite{Essl}], however
the sensitivity of such probe in integrable systems can be very
small.

\section{Acknowledgement}
We would like to thank I. Affleck, E. Altman, I. Bloch, M. Lukin, S.
Lukyanov, J. Schmiedmayer, V. Vuletic  for useful comments and
discussions. This work was partially supported by the NSF,
Harvard-MIT CUA and AFOSR. V.G. was also supported by the Swiss
National Science Foundation, A.P. was supported by AFOSR YIP.


\begin{thebibliography}{99}
\bibitem{footnote} We note that the Luttinger parameter $K$ in coupled
condensates can be additionally renormalized by the tunneling term
and deviate from the single-condensate value. However, we anticipate
that within the Lieb-Liniger model $\beta^2$ is always smaller than
$4\pi$.
\bibitem{quasicl}
We expect that these quasiclassical considerations can be
qualitatively applicable in the large $K$ limit.
\end{thebibliography}
\end{document}